\documentclass[%
reprint,
groupedaddress,
showpacs,
 amsmath,amssymb,
 aps,
prb,
]{revtex4-1}

\newcommand\beq{\begin{equation}}
\newcommand\eeq{\end{equation}}

\usepackage{graphicx}
\usepackage{dcolumn}
\usepackage{bm}

\begin{document}


\title{Tunneling of Obliquely-Incident Waves through \boldmath ${\cal PT}$-Symmetric Epsilon-Near-Zero Bi-Layers}

\author{Silvio Savoia}
\author{Giuseppe Castaldi}
\author{Vincenzo Galdi}
\email{vgaldi@unisannio.it}
\affiliation{Waves Group, Department of Engineering, University of Sannio, I-82100 Benevento, Italy
}%

\author{Andrea Al\`u}
\affiliation{Department of Electrical and Computer Engineering, The University of Texas at Austin, Austin, TX 78712, USA}%

\author{Nader Engheta}
\affiliation{Department of Electrical and Systems Engineering, University of Pennsylvania, Philadelphia, PA 19104, USA}%

\date{\today}


\begin{abstract}
We show that obliquely-incident, transversely-magnetic-polarized plane waves can be totally transmitted (with zero reflection) through {\em epsilon-near-zero} (ENZ) bi-layers characterized by balanced loss and gain with {\em parity-time} (${\cal PT}$) symmetry. This tunneling phenomenon is mediated by the excitation of a surface-wave localized at the interface separating the loss and gain regions. We determine the parameter configurations 
for which the phenomenon may occur and, in particular, the relationship between the incidence direction and the electrical thickness. We show that, below a critical threshold of gain and loss, there always exists a tunneling angle which, for moderately thick (wavelength-sized) structures, approaches a critical value dictated by the surface-wave phase-matching condition.  We also investigate the {\em unidirectional} character of the tunneling phenomenon, as well as the possible onset of {\em spontaneous symmetry breaking}, typical of ${\cal PT}$-symmetric systems. Our results constitute an interesting example of a ${\cal PT}$-symmetry-induced tunneling phenomenon, and may open up intriguing venues in the applications of ENZ materials featuring loss and gain.
\end{abstract}


\maketitle

\section{Introduction}

In a series of seminal works by Bender and co-workers,\cite{Bender:1998,Bender:1999,Bender:2007kr} it was shown that, in spite of the standard axioms in quantum mechanics, a {\em non-Hermitian} Hamiltonian characterized by the so-called {\em parity-time} ($\cal{PT}$) symmetry can still exhibit an {\em entirely real} energy eigenspectrum. Similar concepts had also been previously explored within the framework of atomic physics. \cite{Oberthaler:1996} 

In essence, a $\cal{PT}$-symmetric Hamiltonian commutes with  the combined {\em parity} (i.e., space-reflection, ${\bf r}\rightarrow -{\bf r}$) and {\em time-reversal} ($t\rightarrow -t$, or complex-conjugation $^*$ in the time-harmonic regime) operator.\cite{Bender:2007kr}
This implies that the quantum potential satisfies the symmetry condition $V({\bf r})=V^*(-{\bf r})$. 
However, the latter is only a {\em necessary} condition for the so-called ``exact'' phase characterized by a {\em real} eigenspectrum. Beyond some non-Hermiticity threshold, 
an abrupt phase transition may occur to the so-called ``broken'' phase characterized by a {\em complex} eigenspectrum.\cite{Bender:2007kr} 
Such phenomenon, typically referred to as {\em spontaneous symmetry breaking}, may occur since the Hamiltonian and the (anti-linear) ${\cal PT}$ operators do not necessarily share the same eigenstates.\cite{Bender:2007kr}

More recently, the $\cal{PT}$-symmetry concept has elicited a great deal of interest within the fields of optics, photonics, and plasmonics.
Theoretically founded on the formal analogies between Helmholtz and Schr\"odinger equations, such interest is motivated 
by the relatively simpler (by comparison with quantum physics) conception and realization of ${\cal PT}$-symmetric electromagnetic structures by means of spatially-modulated distributions of loss and gain, either across or along the wave-propagation direction. Within this framework, the arguably simplest scenario consists of coupled optical waveguides, either in passive, lossy configurations 
\cite{Guo:2009hd} (which, after appropriate transformations, are pseudo ${\cal PT}$-symmetric) or in the actual presence of loss and gain. \cite{Ruter:2010dy} Starting from these simpler configurations, a number of different ${\cal PT}$-symmetry-inspired phenomena and effects have been studied in optical, plasmonic, and metamaterial structures, from both theoretical/numerical \cite{Mostafazadeh:2009ea,Ctyroky:2010de,Mostafazadeh:2011im,Jones:2011km,Benisty:2011jr,Miroshnichenko:2011ba,Chong:2011ev,Longhi:2011wy,Jones:2012kf,Miri:2012bq,Ge:2012bq,Yu:2012,Alaeian:2013tu,Kang:2013dg,Castaldi:2013ff,Miri:2013gk,Kulishov:2013gs,Luo:2013}
and experimental \cite{Regensburger:2012jm,Feng:2012jj,Regensburger:2013kd} sides. Besides the very intriguing application-oriented perspectives in the development of novel devices and components (e.g., switches, lasers, absorbers), optical and photonic analogies may also serve as more feasible experimental testbeds for controversial ${\cal PT}$-symmetry-induced quantum-field effects. \cite{Longhi:2012ws}
Also worth of mention is the research field of ${\cal PT}$-symmetric electronics, based on circuit implementations where gain can naturally be introduced via amplifiers.\cite{Schindler:2012,Lin:2012}

A particularly simple and yet very insightful ${\cal PT}$-symmetric optical configuration is obtained by pairing two material slabs characterized by permittivities (and/or permeabilities) with same real part and opposite imaginary parts, i.e., loss at one side and gain at the other side. In the topical literature, such ${\cal PT}$-symmetric bi-layers have been studied extensively under normally-incident plane-wave illumination, showing intriguing anomalous effects such as spectral singularities,\cite{Mostafazadeh:2009ea}  coherent perfect absorption,\cite{Chong:2011ev} and anisotropic transmission resonances.\cite{Ge:2012bq} 
Against this background, in this paper, we deal instead with {\em oblique} plane-wave illumination. We show that, for suitable field polarization and constitutive parameters, a propagating plane-wave obliquely impinging from vacuum  may effectively tunnel (with zero reflection) through the ${\cal PT}$-symmetric bi-layer. This phenomenon is mediated by the excitation of a localized surface wave at the interface separating the gain and loss regions, and generally exhibits a {\em unidirectional} character, i.e., zero reflection is achieved only when exciting the structure from one side and not from the other. 

In particular, we focus on the {\em epsilon-near-zero} (ENZ) regime, i.e., vanishingly small real part of the permittivities, for which the above phenomena may be observed even in the presence of reasonably low levels of loss and gain, and moderately thick structures. 
Recently, ENZ materials have gained a growing attention, \cite{Engheta:2013il} and their application has been suggested in a variety of scenarios including, among others, supercoupling, \cite{Silveirinha:2006ed}
tailoring the radiation phase pattern of sources,\cite{Alu:2007bm}
dielectric sensing,\cite{Alu:2008gm} enhancing the photon density of state for embedded emitters, \cite{Alu:2009ku,Vesseur:2013bs}
 boosting nonlinear effects, \cite{Rizza:2011we,Argyropoulos:2012bu,Vincenti:2013rd} subwavelength image manipulation, \cite{Castaldi:2012cw} nonlocal transformation optics, \cite{Castaldi:2012jq}, and field enhancement. \cite{Campione:2013bj} 
Moreover, for these materials, the effects of loss and gain have been studied in connection with
loss compensation, \cite{Rizza:2011be,Campione:2011fr,deCeglia:2013ca}
perfect absorption and giant magnification, \cite{Jin:2011kd}
loss-enhanced transmission and collimation, \cite{Sun:2012gq}
coherent-perfect absorption, \cite{Feng:2012}
loss-induced omnidirectional bending, \cite{Feng:2012rt}
and gain-assisted harmonic generation. \cite{Vincenti:2012kk}
Our study here provides a new perspective in the effect of balanced loss and gain in ENZ materials. 

The rest of the paper is organized as follows. After an outline of the problem (geometry, assumptions, and observables) in Sec. \ref{Sec:Problem}, we present the main analytical derivations and numerical results (Sec. \ref{Sec:Results}), with details relegated in three Appendices. In connection with the tunneling condition, we identify three regimes of operation (depending on the loss/gain level), and the possible presence of a critical tunneling angle dictated by the dispersion law of the surface-wave excited at the interface separating the loss and gain regions. We also study the {\em unidirectional} character of the phenomenon, as well as the onset of spontaneous symmetry-breaking (Sec. \ref{Sec:SSB}) and, finally, provide some conclusions and perspectives (Sec. \ref{Sec:Conclusions}).

%
\begin{figure}
\begin{center}
\includegraphics [width=8.5cm]{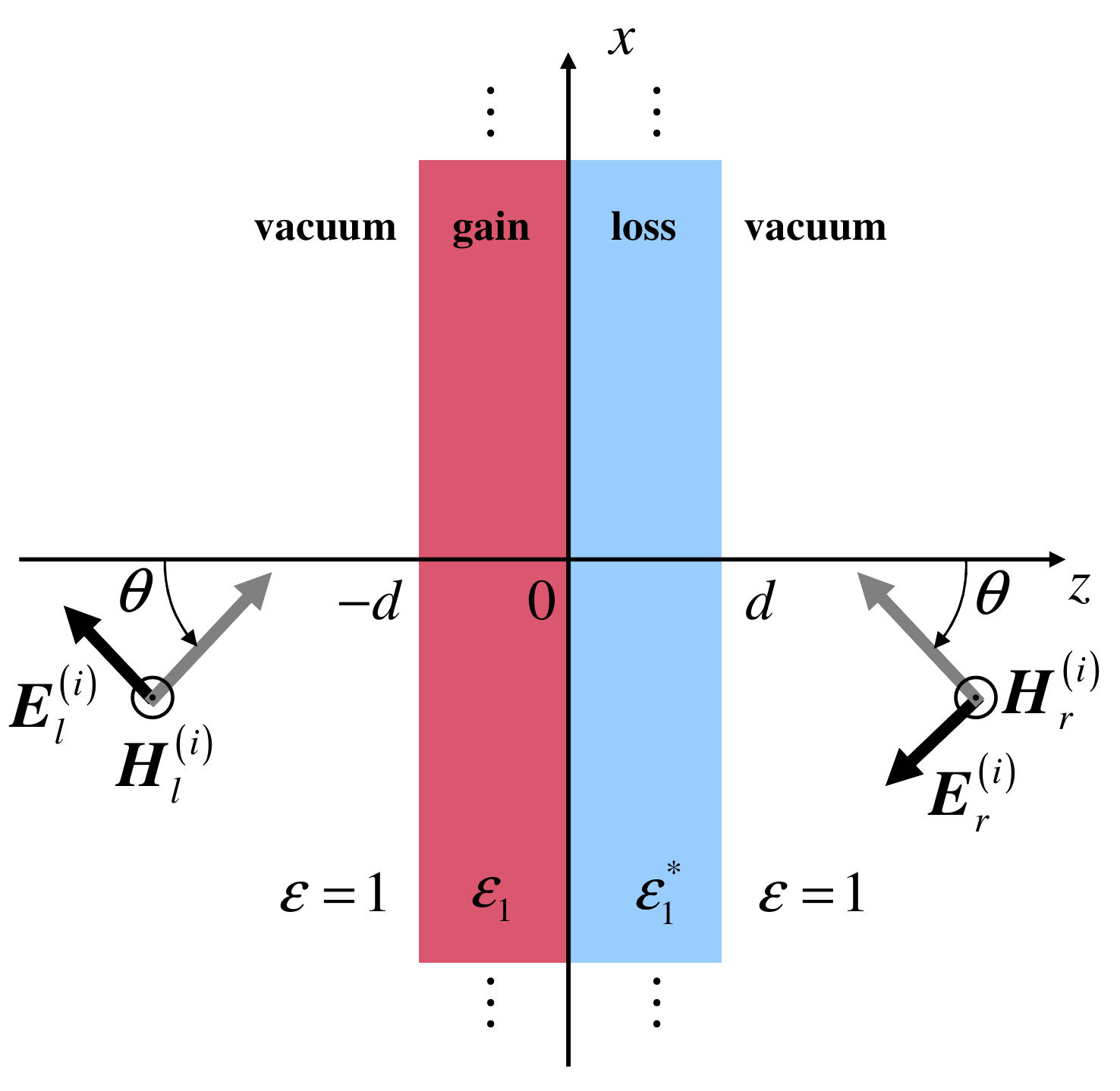}
\end{center}
\caption{(Color online) Problem geometry. A ${\cal PT}$-symmetric bi-layer composed of two halves of identical thickness $d$, and relative permittivities $\varepsilon_1=\varepsilon'-i\varepsilon''$ ($\varepsilon''>0$, i.e., gain) and $\varepsilon_1^*=\varepsilon'+i\varepsilon''$ (loss), immersed in vacuum, is illuminated by a TM-polarized plane wave obliquely impinging from the left or right side.}
\label{Figure1}
\end{figure}

\section{Problem Statement}
\label{Sec:Problem}

\subsection{Geometry and Assumptions}
\label{Sec:Geom}
As illustrated in Fig. \ref{Figure1}, the geometry of interest features an isotropic, non-magnetic (i.e., relative permeability $\mu=1$), piece-wise homogeneous ${\cal PT}$-symmetric bi-layer immersed in vacuum. The bi-layer is composed of two slabs of identical thickness $d$ (and infinite extent along the $x,y$ directions) paired along the $z$-direction, characterized by complex-conjugate relative permittivities $\varepsilon_1$ and $\varepsilon_1^*$, respectively, so as to fulfill the necessary condition for ${\cal PT}$ symmetry, $\varepsilon(z)=\varepsilon^*(-z)$. Under time-harmonic [$\exp(-i\omega t)$] time-convention, we assume 
\beq
\varepsilon_1=\varepsilon'-i\varepsilon'',~~\varepsilon'>0,~~\varepsilon''>0,
\label{eq:eps1}
\eeq
so that the left and right halves ($-d<z<0$ and $0<z<d$, respectively) are characterized by gain and loss, respectively. Moreover, we focus on the ENZ limit
\beq
\varepsilon'\ll \varepsilon''\ll 1.
\label{eq:ENZ}
\eeq
We are interested in studying the electromagnetic response under transverse-magnetic (TM) plane-wave illumination obliquely-incident from either sides (cf. Fig. \ref{Figure1}). Accordingly, we consider a unit-amplitude, $y$-directed magnetic field ($H^{(i)}_l$ or $H^{(i)}_r$)
\beq
H_{l,r}^{(i)}\left(x,z\right)= \exp\left[i\left(k_{x0}x\pm k_{z0}z\right)\right],
\eeq
where the subscripts $l$ and $r$ (and $+$ and $-$ signs) denote the incidence from the left and right, respectively. For propagating waves with incidence angle $\theta$ (cf. Fig. \ref{Figure1}), the wavenumbers $k_{x0}$ and $k_{z0}$ can be expressed as
\beq
k_{x0}=k_0\sin\theta,~~~k_{z0}=k_0\cos\theta,
\label{eq:kxz0}
\eeq
in terms of the vacuum wavenumber $k_0=\omega/c_0=2\pi/\lambda_0$ (with $c_0$ denoting the speed of light in vacuum, and $\lambda_0$ the corresponding wavelength).

We note that for $\varepsilon''=0$ our scenario reduces to the case already studied in Ref. \onlinecite{Alu:2007bm}, for which wave tunneling was observed at a Brewster angle $\theta_B$ corresponding to the polaritonic resonance of the slab
\beq
k_{x0}=k_0\sin\theta_B=k_0\sqrt{\varepsilon'},
\label{eq:PR}
\eeq
with angular bandwidth narrowing down for decreasing values of $\varepsilon'$ and/or increasing values of $k_0d$. The reader is also referred to Ref. \onlinecite{Lu:2013kw,Luo:2013aw} for other examples of total-transmission through {\em anisotropic}, lossless ENZ slabs. On the other hand, in ENZ slabs characterized by slight (e.g., partially compensated) losses, a pseudo-Brewster angle can be identified for which 
reflection is strongly reduced due to a nonresonant impedance-matching
condition.\cite{deCeglia:2013ca}

Our study below complements the above results, by identifying a different wave-tunneling phenomenon that can occur in ENZ bi-layers with {\em balanced} loss and gain.

\subsection{Observables}
Referring to the incident fields in Fig. \ref{Figure1}, and labeling with the superscripts $^{(R)}$ and $^{(T)}$ the corresponding reflected and transmitted fields, respectively, we define as meaningful observables the reflection coefficients for incidence from the left and right,
\beq
R_l\equiv\frac{H^{(R)}_l\left(x,-d\right)}{H^{(i)}_l\left(x,-d\right)},~~~R_r\equiv\frac{H^{(R)}_r\left(x,d\right)}{H^{(i)}_r\left(x,d\right)},
\label{eq:RRlr}
\eeq
which are generally different, and a transmission coefficient,
\beq
T\equiv\frac{H^{(T)}_l\left(x,d\right)}{H^{(i)}_l\left(x,-d\right)}=
\frac{H^{(T)}_r\left(x,-d\right)}{H^{(i)}_r\left(x,d\right)},
\label{eq:TT}
\eeq
which is identical for both types of incidence, due to reciprocity. The above observables are related by a generalized unitarity relation,\cite{Ge:2012bq}
\beq
R_l R_r=T^2\left(1-\frac{1}{\left|T\right|^2}\right),
\eeq
which, in turn, yields the conservation relation\cite{Ge:2012bq}
\beq
\left|
\left|T
\right|^2-1
\right|=\left|R_l R_r\right|.
\label{eq:unit2}
\eeq

\section{Main Analytical and Numerical Results}

\label{Sec:Results}

\subsection{Reflection and Transmission Coefficients}
The observables defined in (\ref{eq:RRlr}) and (\ref{eq:TT}) can be calculated analytically (see Appendix \ref{Sec:AppA} for details). 
For the reflection coefficients, we obtain
\begin{eqnarray}
R_{l,r}\left(\theta,k_0d,\varepsilon_1\right)\!\equiv\!\frac{N_1\left(\theta,k_0d,\varepsilon_1\right)\!\pm \!N_2\left(\theta,k_0d,\varepsilon_1\right)}
{D\left(\theta,k_0d,\varepsilon_1\right)}\!,
\label{eq:RLR}
\end{eqnarray}
where the $+$ and $-$ signs refer to the incidence from left (i.e., $R_l$) and right (i.e., $R_r$), respectively, and
\begin{subequations}
\begin{eqnarray}
N_1\left(\theta,k_0d,\varepsilon_1\right)&=&\left|k_{z1}\right|^2 \mbox{Re}\left(
\varepsilon_1^* k_{z1}\tau_1\right)\nonumber\\
&-&\left|\varepsilon_1\right|^2 k_{z0}^2 \mbox{Re}\left(\varepsilon_1 k_{z1}^*\tau_1\right),
\label{eq:N1}\\
N_2\left(\theta,k_0d,\varepsilon_1\right)&=&k_{z0} \left|\tau_1\right|^2 \mbox{Re}\left[i\varepsilon_1^2\left(k_{z1}^*\right)^2
\right]
\label{eq:N2}
\nonumber\\
&=&\varepsilon''k_0^3 \left|\tau_1\right|^2 \cos\theta\nonumber\\
&\times&
\left[
\left(\varepsilon'\right)^2+
\left(\varepsilon''\right)^2
-2\varepsilon'\sin^2\theta
\right],
\end{eqnarray}
\label{eq:N123}
\end{subequations}
\begin{eqnarray}
D\left(\theta,k_0d,\varepsilon_1\right)&=&i k_{z0} \left\{\left|\tau_1\right|^2\mbox{Re}\left[\varepsilon_1^2\left(k_{z1}^*\right)^2\right]
-\left|\varepsilon_1\right|^2 \left|k_{z1}\right|^2
\right\}\nonumber\\
&-&\left|k_{z1}\right|^2\mbox{Re}\left(\varepsilon_1k_{z1}^*\tau_1^*\right)\nonumber\\
&-&\mbox{Re}\left(\left|\varepsilon_1\right|^2\varepsilon_1 k_{z0}^2k_{z1}^*\tau_1\right),
\end{eqnarray}
with 
\beq
k_{z1}=k_0\sqrt{\varepsilon_1-\sin^2\theta},~~\mbox{Im}\left(k_{z1}\right)\le0,
\label{eq:kz1}
\eeq
\beq
\tau_1=\tan\left(k_{z1}d\right).
\label{eq:tau1}
\eeq
The transmission coefficient can be instead written as
\beq
T\left(\theta,k_0d,\varepsilon_1\right)=\frac{-i k_{z0} \left|\varepsilon_1\right|^2 \left|k_{z1}\right|^2 \left|1+\tau_1^2\right|}
{D\left(\theta,k_0d,\varepsilon_1\right)}.
\label{eq:TLR}
\eeq

\subsection{Tunneling Conditions}

\label{Sec:Tunneling}
The tunneling (i.e., zero-reflection) conditions, for incidence from either side, can be derived from (\ref{eq:RLR}) by enforcing
\beq
N_1\left(\theta,k_0d,\varepsilon_1\right)\pm N_2\left(\theta,k_0d,\varepsilon_1\right)=0,
\label{eq:tunnel}
\eeq
subject to {\em a posteriori} verification that the denominator is nonzero.
From (\ref{eq:N2}), it immediately follows that
\beq
N_2\left(\theta_c,k_0d,\varepsilon_1\right)=0,~~~\theta_c=\arcsin\left[
\sqrt{\frac{\left(\varepsilon'\right)^2+\left(\varepsilon''\right)^2}{2\varepsilon'}}
\right].
\label{eq:N2lim}
\eeq
Moreover, it can be shown (see Appendix \ref{Sec:AppB} for details) that
\beq
\lim_{k_0d\rightarrow\infty}N_1\left(\theta_c,k_0d,\varepsilon_1\right)=0,
\label{eq:N13lim}
\eeq
which implies that, for a sufficiently thick bi-layer, the tunneling condition (in $\theta$) approaches the critical angle $\theta_c$ in (\ref{eq:N2lim}). Such angle, which admits real values for
\beq
\varepsilon''\le \varepsilon_u \equiv\sqrt{\varepsilon'\left(2-\varepsilon'\right)},
\label{eq:BTR}
\eeq
is fundamentally different from the standard Brewster angle in a conventional (lossless, gainless) dielectric slab
\beq
\theta_B=\arctan{\sqrt{\varepsilon'}},
\label{eq:SBA}
\eeq
and it is also not related to the polaritonic resonance of a lossless ENZ slab in (\ref{eq:PR}) (see also the discussion in Sec. \ref{Sec:ENZ} below). For a better understanding, we observe that the configuration in Fig. \ref{Figure1} may support (in the halfspace limit $d\rightarrow\infty$) a surface wave exponentially bound along the $z$ direction, characterized by the dispersion relationship\cite{Nezhad:2004fg,Ctyroky:2010de}
\beq
k_x^{(SW)}=k_0\sqrt{\frac{\varepsilon_1\varepsilon_1^*}{\varepsilon_1+\varepsilon_1^*}}=k_0\sqrt{\frac{\left(\varepsilon'\right)^2+\left(\varepsilon''\right)^2}{2\varepsilon'}}.
\label{eq:kSW}
\eeq
Interestingly, the inherent ${\cal PT}$-symmetry dictates that the propagation constant in (\ref{eq:kSW}) is {\em always real}. Moreover, it can be observed that the critical angle $\theta_c$ in (\ref{eq:N2lim}) yields the phase-matching condition for the coupling of the impinging plane wave with the surface wave in (\ref{eq:kSW}). 

%
\begin{figure}
\begin{center}
\includegraphics [width=8.5cm]{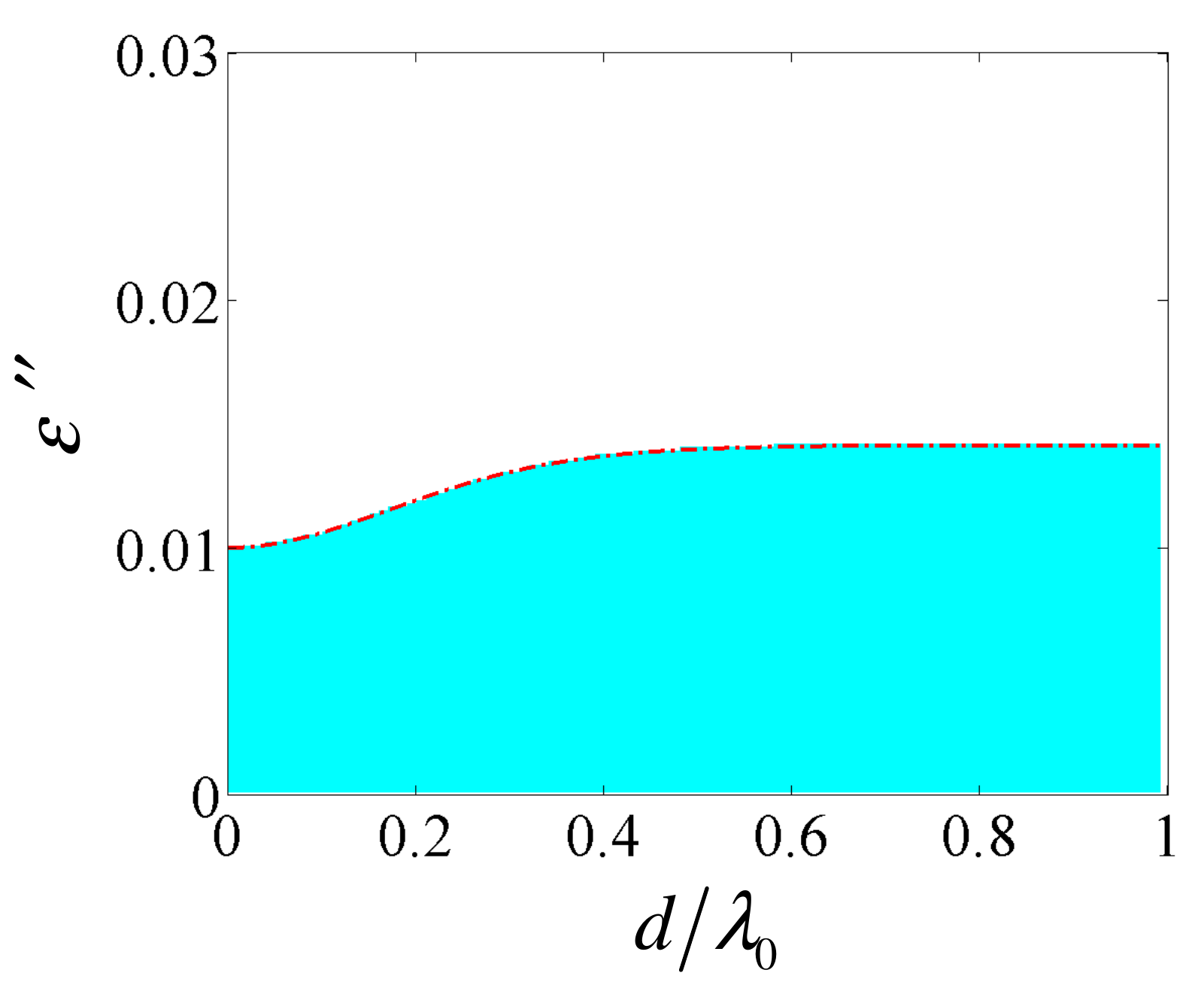}
\end{center}
\caption{(Color online) Configuration as in Fig. \ref{Figure1}, but in the ENZ limit ($\varepsilon'=10^{-4}$). The cyan-shaded area identifies parameter configurations ($d/\lambda_0$, $\varepsilon''$) for which a tunneling angle is numerically found. The dash-dotted curve represents the approximate analytical bound in (\ref{eq:epsbound}).}
\label{Figure2}
\end{figure}

\subsection{Results}
\label{Sec:ENZ}
The above observations imply that, for moderate to large electrical thicknesses, the tunneling phenomenon is mediated by the excitation of a surface wave at the gain-loss interface $z=0$. Nevertheless, the tunneling phenomenon can also be observed in electrically-thin structures. In the ENZ limit, it can be shown (see Appendix \ref{Sec:AppC} for details) that, for given bi-layer permittivities and electrical thickness, there {\em always} exists an incidence angle yielding tunneling (from either side) within the interval $(\theta_c,\pi/2)$, provided that
\beq
\varepsilon''\le\sqrt{
\frac{\varepsilon'\left(2-\varepsilon'\right)\left[\varepsilon' k_0d \left(\tau_0^2-1\right)+2\tau_0\right]}
{k_0d
\left(\varepsilon'-2\right)\left(\tau_0^2-1\right)
+2\tau_0}
},~\tau_0\!=\!\tanh\left(k_0d\right)\!.
\label{eq:epsbound}
\eeq
We note that real solutions of (\ref{eq:epsbound}) exist if
\beq
\varepsilon'<\frac{2\tau_0}{k_0d\left(1-\tau_0^2\right)},
\eeq
which is satisfied in the assumed ENZ limit. Moreover, we observe that, in the limit $k_0d\rightarrow 0$ (i.e., $\tau_0\rightarrow 0$), the bound in (\ref{eq:epsbound}) reduces to 
\beq
\varepsilon''\le \varepsilon''_l\equiv \left(2-\varepsilon'\right)\sqrt{\frac{\varepsilon'}{4-\varepsilon'}},
\eeq
whereas, in the asymptotic limit $k_0d\rightarrow \infty$ (i.e., $\tau_0\rightarrow 1$), it reduces to
the condition in (\ref{eq:BTR}), which ensures real values of the critical angle in (\ref{eq:N2lim}).

For a representative small value of $\varepsilon'$, Fig. \ref{Figure2} illustrates and numerically verifies the above bound via a bi-partition of the relevant parameter space ($\varepsilon''$ vs. $d/\lambda_0$) between a region (cyan shaded) where tunneling angles are numerically found, and another region (white) where no solution can be found. As it can be observed, such numerical bi-partition is in excellent agreement with the approximate estimate in (\ref{eq:epsbound}).

From Fig. \ref{Figure2}, we can basically identify three representative parameter configurations, as detailed below.

%
\begin{figure*}
\begin{center}
\includegraphics [width=16cm]{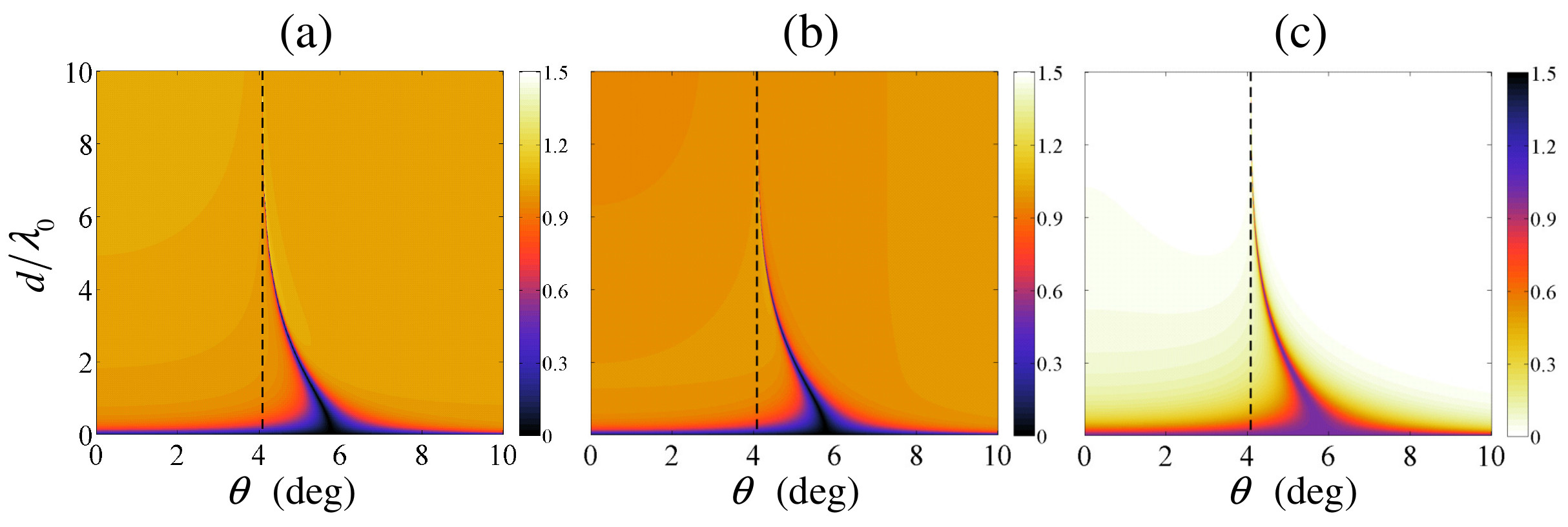}
\end{center}
\caption{(Color online) Observables (magnitude) in (\ref{eq:RLR}) and (\ref{eq:TLR}) as a function of $\theta$ and $d/\lambda_0$, for $\varepsilon'=10^{-4}$, and $\varepsilon''=0.001$ ($\varepsilon''<\varepsilon''_l$). (a) $|R_l|$, (b) $|R_r|$, (c) $|T|$. The vertical dashed lines indicate the critical angle $\theta_c=4.07^o$ [cf. (\ref{eq:N2lim})].}
\label{Figure3}
\end{figure*}

%
\begin{figure}
\begin{center}
\includegraphics [width=8.5cm]{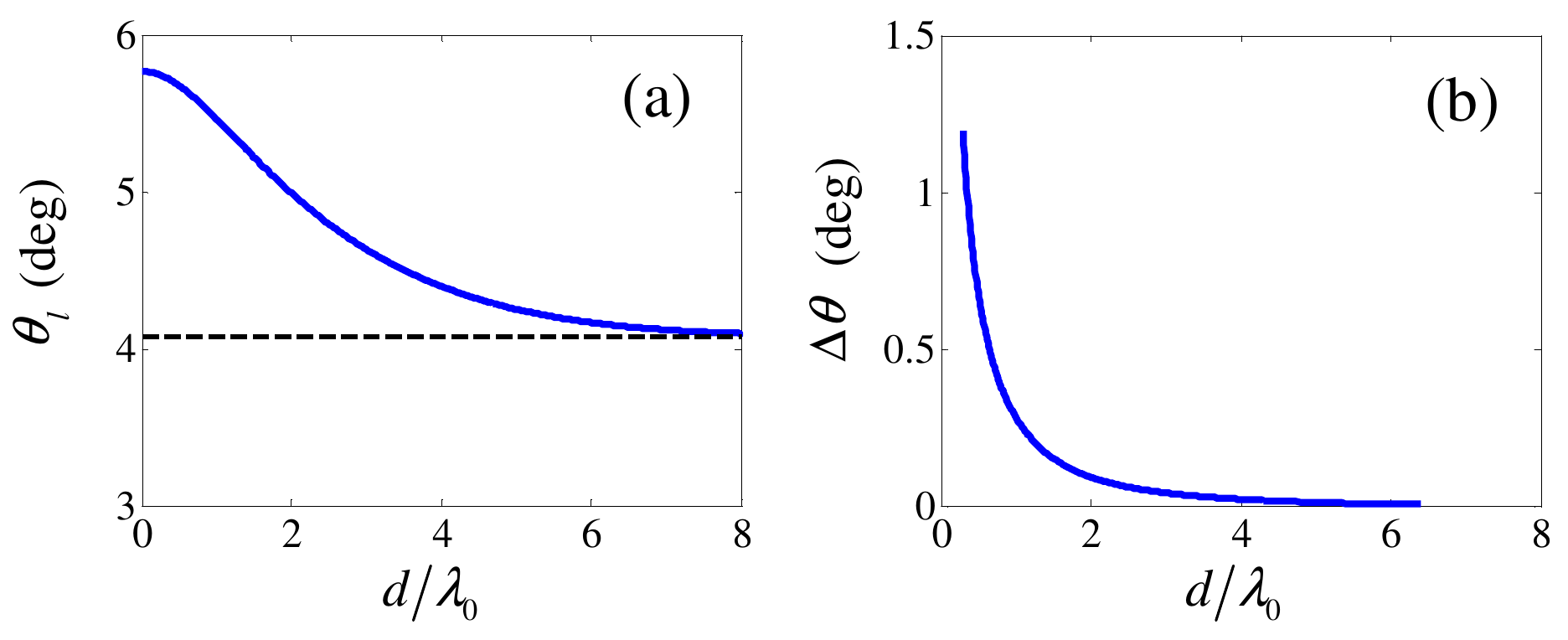}
\end{center}
\caption{(Color online) Parameters as in Fig. \ref{Figure3}. (a) Tunneling angle (for incidence from left) $\theta_l$ as a function $d/\lambda_0$ [extracted from Fig. \ref{Figure3}(a)]. (b) Corresponding full-width at half-maximum $\Delta\theta$ in the transmission response [numerically extracted from Fig. \ref{Figure3}(c)]. The horizontal dashed line indicates the critical angle $\theta_c=4.07^o$ [cf. (\ref{eq:N2lim})].}
\label{Figure4}
\end{figure}

%
\begin{figure*}
\begin{center}
\includegraphics [width=14cm]{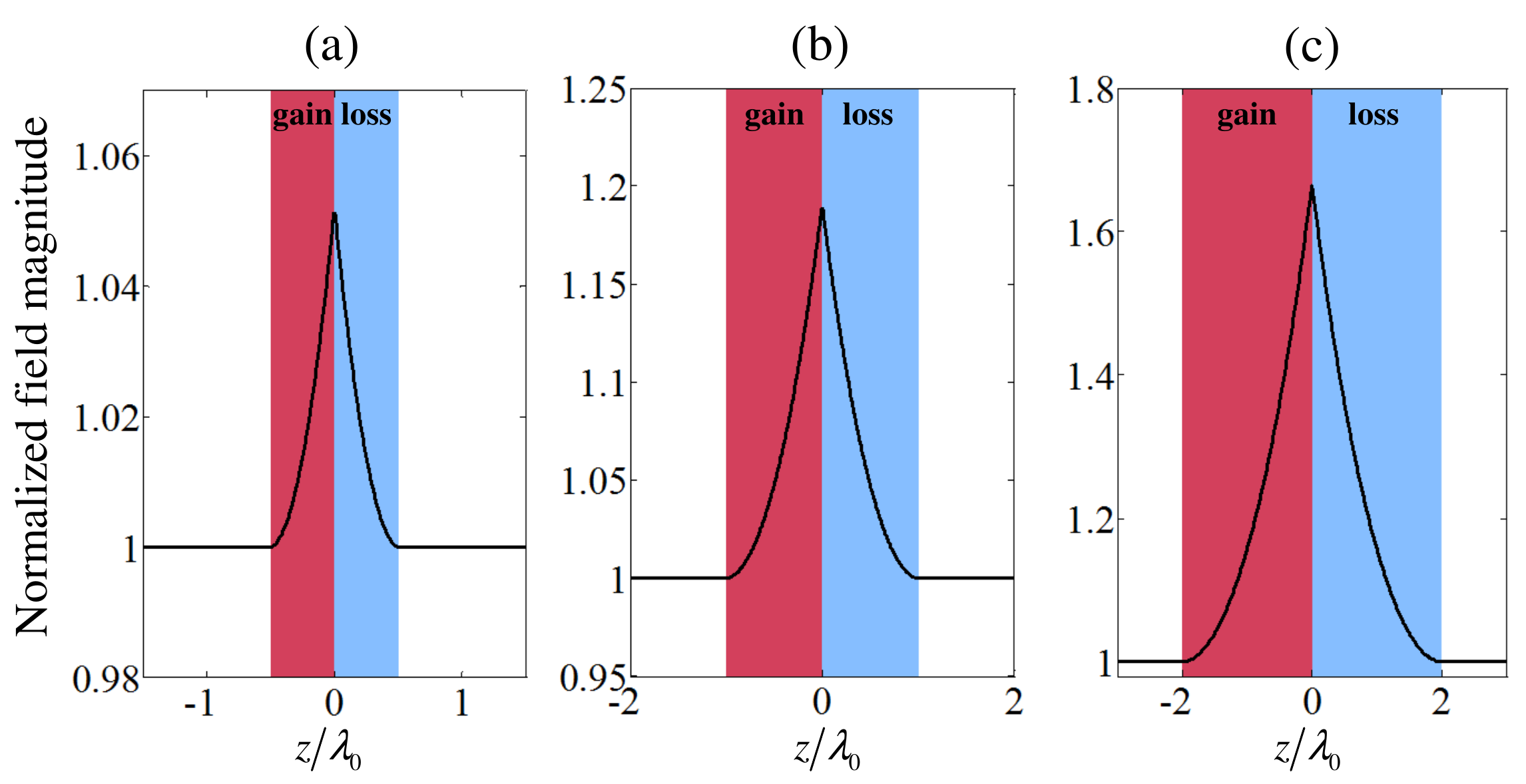}
\end{center}
\caption{(Color online) Parameters as in Fig. \ref{Figure3}. Magnetic-field magnitude ($|H_y|$) distribution (normalized with respect to the incident field) along the $z$-direction, for three representative values of $d/\lambda_0$, and corresponding tunneling angles [for incidence from left, cf. Fig. \ref{Figure4}(a)]. 
(a) $d/\lambda_0=0.5$, $\theta=\theta_l=5.66^o$ 
(b) $d/\lambda_0=1$, $\theta=\theta_l=5.45^o$ 
(c) $d/\lambda_0=2$, $\theta=\theta_l=4.97^o$. 
Note the different scales in the graphs.}
\label{Figure5}
\end{figure*}

%
\begin{figure*}
\begin{center}
\includegraphics [width=16cm]{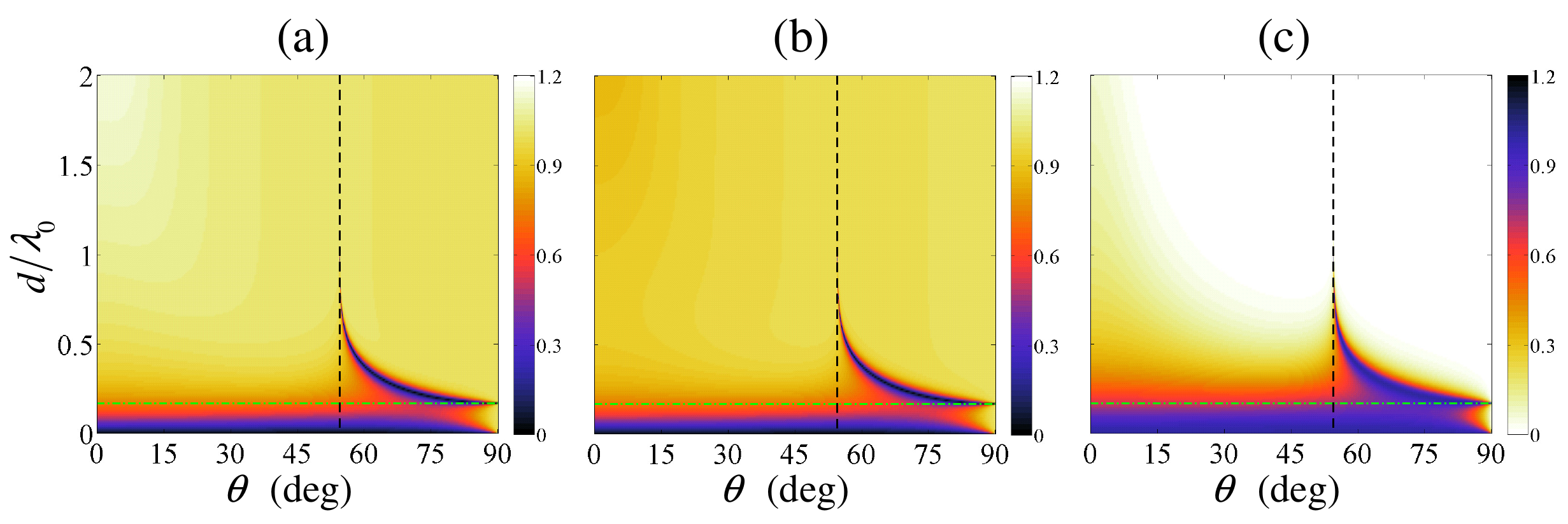}
\end{center}
\caption{(Color online) As in Fig. \ref{Figure3}, but for $\varepsilon''=0.0115$. The horizontal dashed line indicates the critical value of $d/\lambda_0$ beyond which a tunneling angle should exist, according to the approximate estimate in (\ref{eq:epsbound}).}
\label{Figure6}
\end{figure*}

%
\begin{figure}
\begin{center}
\includegraphics [width=8.5cm]{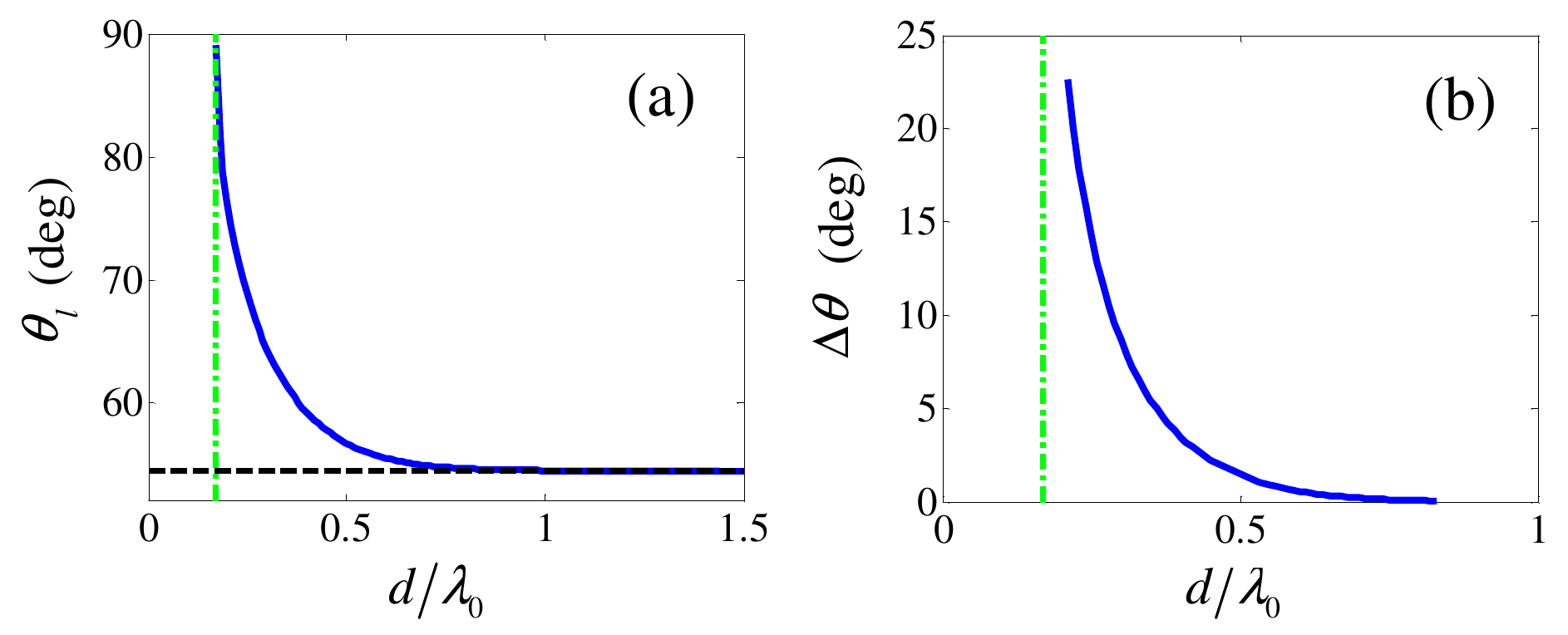}
\end{center}
\caption{(Color online) As in Fig. \ref{Figure4}, but for $\varepsilon''=0.0115$. 
The horizontal dashed line indicates the critical angle $\theta_c=54.41^o$ [cf. (\ref{eq:N2lim})].
The vertical dash-dotted line indicate the critical value $d/\lambda_0=0.17\lambda_0$ [cf. Fig. \ref{Figure2}] beyond which a tunneling angle should exist, according to the approximate estimate in (\ref{eq:epsbound}).}
\label{Figure7}
\end{figure}

%
\begin{figure*}
\begin{center}
\includegraphics [width=16cm]{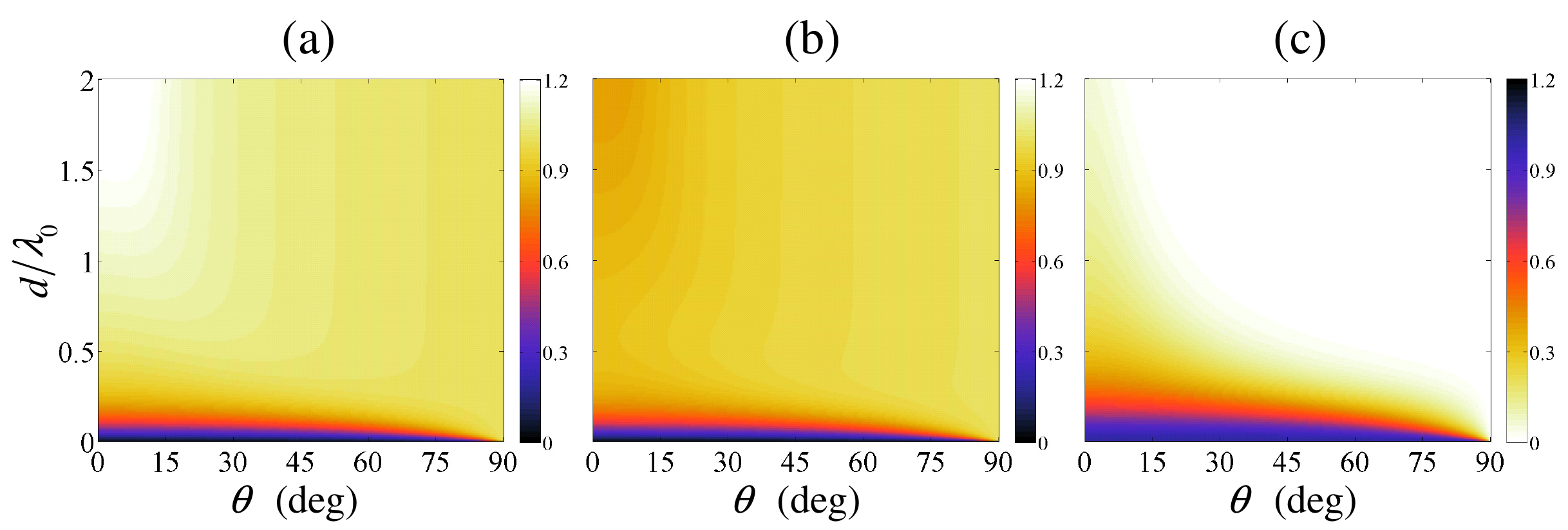}
\end{center}
\caption{(Color online) As in Fig. \ref{Figure3}, but for $\varepsilon''=0.02$.}
\label{Figure8}
\end{figure*}

\subsubsection{$\varepsilon''<\varepsilon''_l$}
In this case, it is always possible to achieve tunneling with a real incidence angle $\theta_c<\theta<\pi/2$, for arbitrary values of the bi-layer electrical thickness. This is illustrated in Fig. \ref{Figure3}, which shows the reflection (from both sides) and transmission coefficient magnitudes, as a function of $\theta$ and $d/\lambda_0$. As predicted, it can be observed that, for any value of $d/\lambda_0$, and for either incidence sides, there always exists a tunneling angle. We note that, though quite similar, the responses for incidence from left ($|R_l|$) and right ($|R_r|$) are actually slightly different, so that a tunneling condition for incidence from left generally implies {\em non-zero reflection} for incidence from right; such {\em unidirectional} character will be discussed in more detail in Sec. \ref{Sec:SSB} below. Moreover, we observe that, for increasing values of the electrical thickness, the reflection dips (and corresponding transmission peaks) gradually move from $\pi/2$ and asymptotically approach the critical angle $\theta_c$ in (\ref{eq:N2lim}), becoming increasingly narrower. This is better quantified in Fig. \ref{Figure4}, which shows the tunneling angle (for incidence from left) $\theta_l$ [numerically extracted from Fig. \ref{Figure3}(a)] and corresponding full-width-half-maximum $\Delta\theta$ in the transmission response [numerically extracted from Fig. \ref{Figure3}(c)] as a function of $d/\lambda_0$. The above results [together with (\ref{eq:N2lim})] allow, in principle, to engineer the phenomenon (in terms of tunneling direction and angular bandwidth) by acting on the bi-layer electrical thickness and constitutive parameters.

For the same parameter configuration, Fig. \ref{Figure5} shows the magnetic-field distributions (along the $z$-direction) corresponding to a tunneling condition for incidence from left, for three representative values of $d/\lambda_0$. The localized field distribution peaked at the interface $z=0$ and field enhancement (increasing with increasing thickness values) confirm that the tunneling phenomenon is mediated by the excitation of a surface wave [cf. (\ref{eq:kSW})]. 

We note that the parameter range of interest ($\varepsilon''<\varepsilon''_l$) also includes the lossless (and gainless) limit $\varepsilon''\rightarrow 0$, which was already studied in Ref. \onlinecite{Alu:2007bm}. In this regime, recalling (\ref{eq:N2}), the tunneling condition in (\ref{eq:tunnel}) becomes
\beq
N_1\left(\theta,k_0d,\varepsilon_1\right)=0,
\label{eq:tunnel1}
\eeq
which is obviously independent on the incidence side, and it can be shown to trivially reduce to the polaritonic resonance condition in (\ref{eq:PR}).

\subsubsection{$\varepsilon''_l<\varepsilon''<\varepsilon''_u$}
In this case, from Fig. \ref{Figure2}, we expect that tunneling angles exist only for values of $d/\lambda_0$ above a critical threshold. This is confirmed by the results in Figs. \ref{Figure6} and \ref{Figure7}, which qualitatively differ from those in Figs. \ref{Figure3} and \ref{Figure4}, respectively, only in the small-thickness region, wherein the reflection never vanishes. For larger values of $d/\lambda_0$, the behavior qualitatively resembles that observed in the previous example, with the reflection dips asymptotically approaching the critical angle $\theta_c$ in (\ref{eq:N2lim}) and narrowing down. However, by comparison with the previous example, this asymptotic regime is approached for smaller values of the electrical thickness (note the different $d/\lambda_0$ scales). In this regime, the field distributions (not shown for brevity) qualitatively resemble those in Fig. \ref{Figure5}.

\subsubsection{$\varepsilon''>\varepsilon''_u$}
In this regime, tunneling conditions are no longer achievable. From the physical viewpoint, recalling  (\ref{eq:N2lim}) and (\ref{eq:kSW}), this is due to the impossibility of exciting a surface wave at the interface $z=0$ with a propagating plane wave impinging from vacuum, i.e., with real values of the critical angle $\theta_c$. This is illustrated in Fig. \ref{Figure8}, which is markedly different from Figs. \ref{Figure3} and \ref{Figure6} above.

%
\begin{figure*}
\begin{center}
\includegraphics [width=16cm]{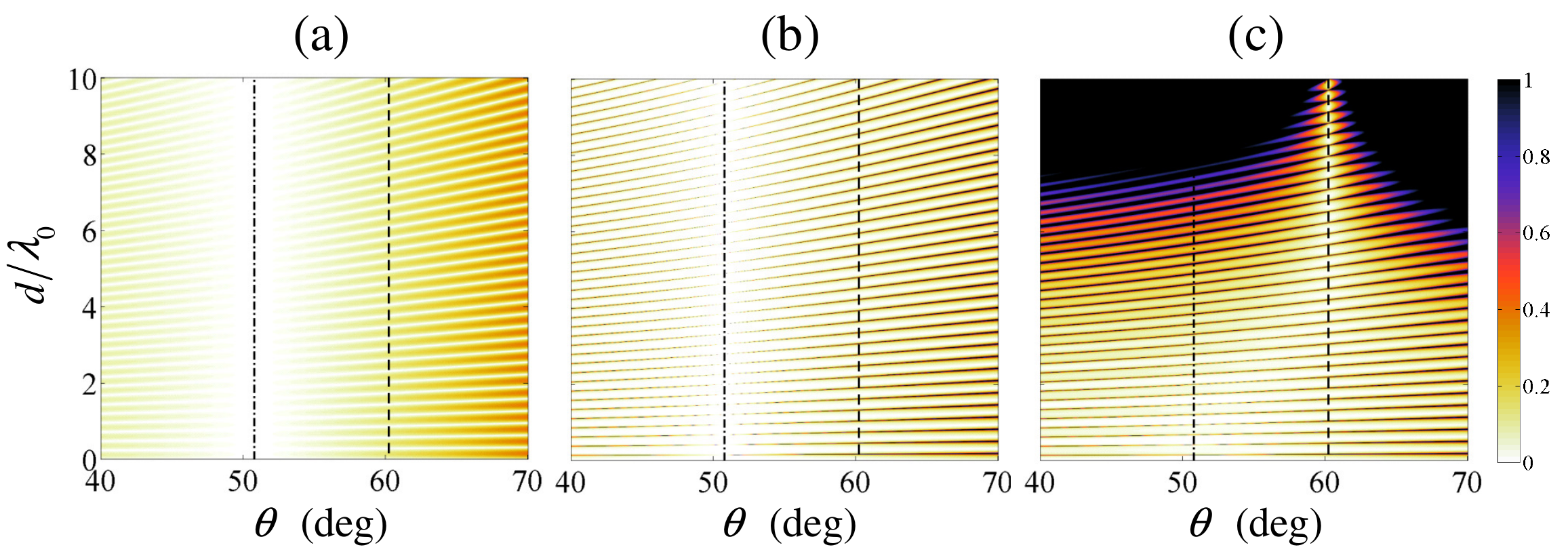}
\end{center}
\caption{(Color online) Reflection coefficient magnitude (for incidence from left) $|R_l|$ as a function of the incidence angle $\theta$ and the bi-layer electrical (semi)thickness $d/\lambda_0$, for  $\varepsilon'=1.5$ and three representative values of $\varepsilon''$. (a) $\varepsilon''=0$, (b) $\varepsilon''=0.001$, (c) $\varepsilon''=0.1$. The vertical dashed and dash-dotted lines indicate the critical angle $\theta_c=60.22^o$ [cf. (\ref{eq:N2lim})] and the Brewster angle $\theta_B=50.77^o$ [cf. (\ref{eq:SBA})] pertaining to a dielectric slab of relative permittivity $\varepsilon'$, respectively.}
\label{Figure9}
\end{figure*}

\subsection{Remarks}
We highlight that the possible excitation of a surface wave at the gain-loss interface $z=0$, and hence the associated tunneling phenomenon, is not necessarily restricted to the insofar considered ENZ regime. From the mathematical viewpoint, the existence of a real-valued critical angle $\theta_c$ in (\ref{eq:N2lim}) can be guaranteed by the condition $\varepsilon'< 2$ [cf. (\ref{eq:BTR})]. However, it can be observed from (\ref{eq:tunnel}) and (\ref{eq:N2}) that, for moderate values of $\varepsilon'$ and $k_0d$, this mechanism becomes effectively dominant only for values of $\varepsilon''$ corresponding to unfeasibly high levels of gain. This is exemplified in Fig. \ref{Figure9}, which 
shows the reflection-coefficient magnitude for incidence from left (i.e., $|R_l|$), as a function of the incidence angle and electrical thickness, for a ${\cal PT}$-symmetric bi-layer with $\varepsilon'=1.5$ and three representative values of $\varepsilon''$. In particular, the reference case $\varepsilon''=0$ (i.e., no loss and gain) in Fig. \ref{Figure9}(a) illustrates the standard Fabry-Perot-type oscillations, as well as the standard Brewster-angle condition
typical of dielectric slabs for oblique, TM illumination. Very similar results can be observed for a configuration featuring low levels of loss and gain [$\varepsilon''=0.001$, cf. Fig. \ref{Figure9}(b)]. By further increasing $\varepsilon''$ up to high
levels of gain [$\varepsilon''=0.1$ cf. Fig. \ref{Figure9}(c)], the tunneling phenomenon becomes barely visible, with the zero-reflection ridge disappearing at the standard Brewster-angle $\theta_B=56.3^o$ in (\ref{eq:SBA}), and gradually appearing (beyond a critical thickness value) at the critical angle $\theta_c=60.22^o$ given by (\ref{eq:N2lim}). In order to obtain a markedly visible phenomenon (as in Fig. 3), one would need unfeasibly high levels of gain. These observations motivate our focus on the ENZ regime, for which the tunneling phenomenon is attainable even in the presence of moderate-to-small electrical thicknesses and low levels of gain and loss.

%
\begin{figure*}
\begin{center}
\includegraphics [width=16cm]{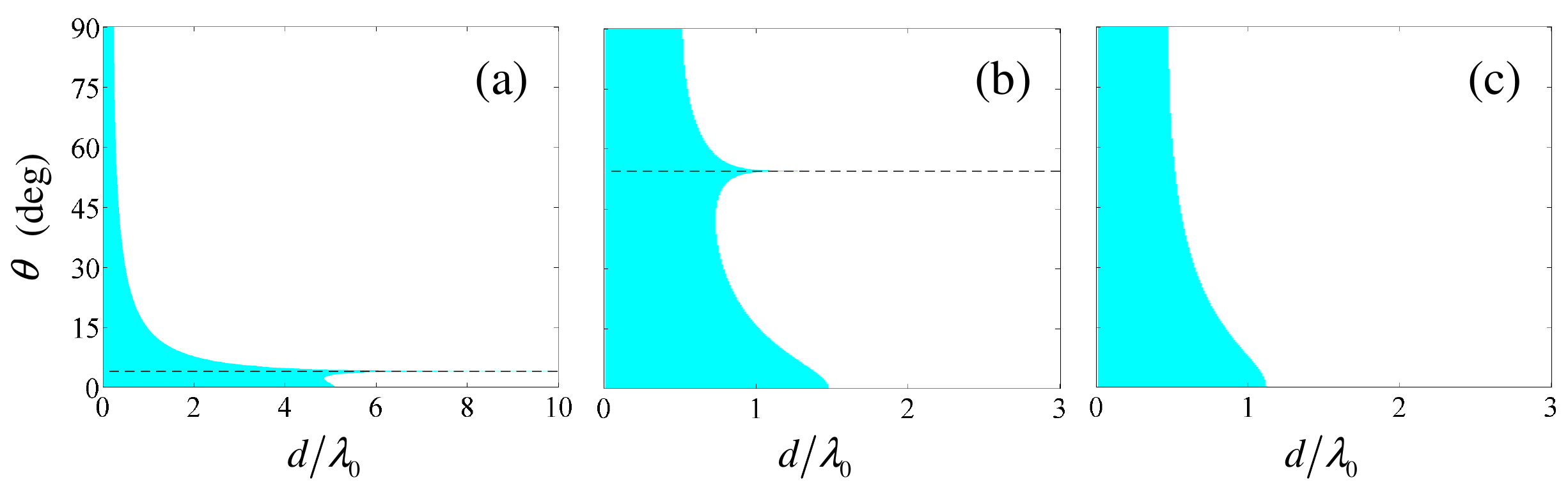}
\end{center}
\caption{(Color online) Illustration of the spontaneous symmetry breaking phenomenon, for $\varepsilon'=10^{-4}$, and representative values of $\varepsilon''$. The cyan-shaded regions indicate the parameters configurations ($d/\lambda_0$, $\theta$) for which the eigenvalues of the scattering matrix ${\underline {\underline S}}_0$ in (\ref{eq:SS}) are unimodular ($|\sigma_1|=|\sigma_2|=1$), as a function of $\theta$ and $d/\lambda_0$,  (a) $\varepsilon''=0.001$, (b) $\varepsilon''=0.0115$, (c) $\varepsilon''=0.02$. The horizontal dashed line indicates the critical angle $\theta_c$ [cf. (\ref{eq:N2lim})]
}
\label{Figure10}
\end{figure*}

%
\begin{figure*}
\begin{center}
\includegraphics [width=16cm]{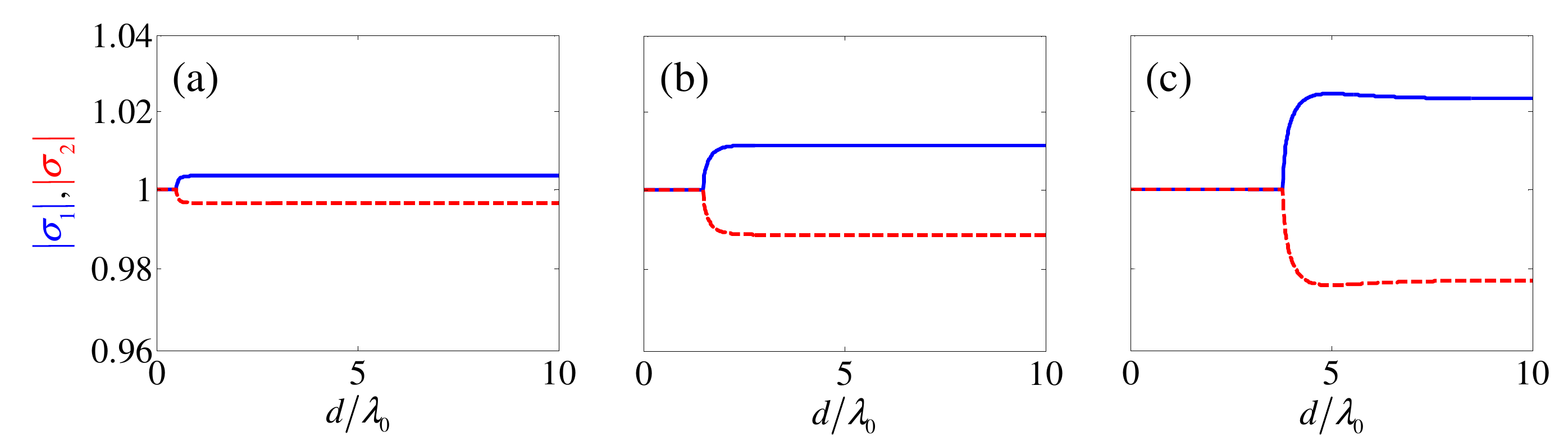}
\end{center}
\caption{(Color online) As in Fig. \ref{Figure10}(a) ($\varepsilon''=0.001$), but magnitude (in semi-log scale) of eigenvalues $\sigma_1$ (blue-solid) and $\sigma_2$ (red-dashed) as a function of $d/\lambda_0$, for three representative incidence angle. (a) $\theta=30^o$, (b) $\theta=10^o$, (c) $\theta=5^o$.}
\label{Figure11}
\end{figure*}

%
\begin{figure}
\begin{center}
\includegraphics [width=8.5cm]{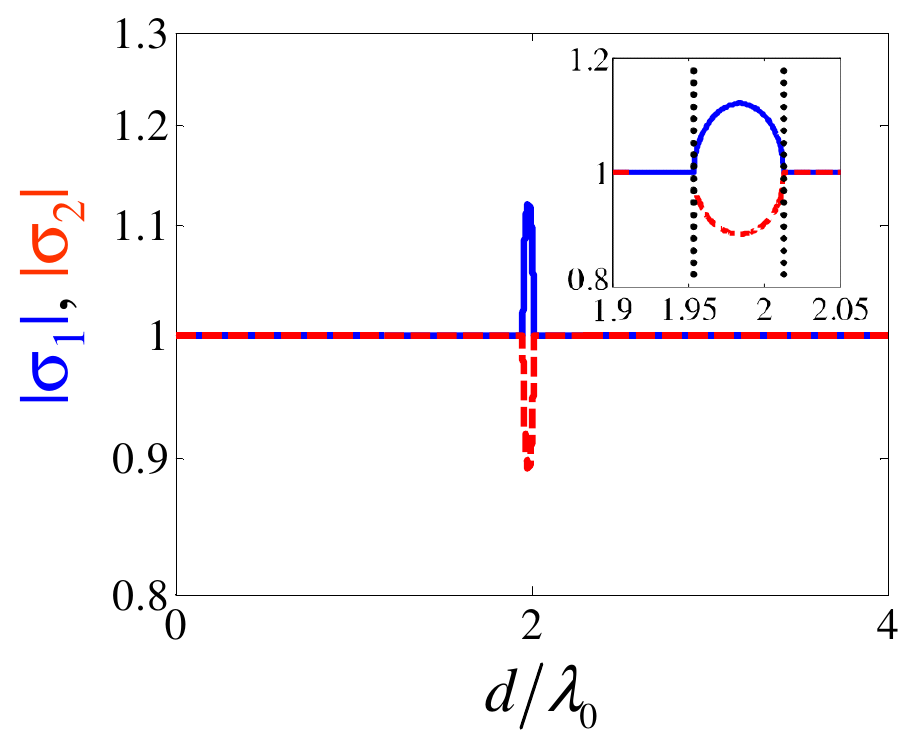}
\end{center}
\caption{(Color online) As in Fig. \ref{Figure11} ($\varepsilon''=0.001$), but eigenvalues of the scattering matrix ${\underline {\underline S}}_c$ in (\ref{eq:SSc}), for $\theta=5^o$. The inset shows a magnified detail around the two exceptional points, which correspond to the  tunneling conditions (vertical dotted lines) for incidence from the left ($R_l=0$, at $d/\lambda_0=1.95$) and right ($R_r=0$, at $d/\lambda_0=2.01$).}
\label{Figure12}
\end{figure}

\section{Spontaneous Symmetry Breaking}
\label{Sec:SSB}
As previously mentioned, the symmetry condition $\varepsilon(z)=\varepsilon^*(-z)$ exhibited by the bi-layer in Fig. \ref{Figure1} is only a necessary, but not sufficient, condition for the eigenspectrum to be real.\citep{Bender:2007kr} For given frequency and incidence direction, beyond a critical threshold of loss/gain level, the so-called ``spontaneous symmetry breaking'' may occur, i.e., 
an abrupt phase transition to a {\em complex} eigenspectrum.\citep{Bender:2007kr} 

In what follows, we investigate this phenomenon by utilizing a standard approach, already applied successfully to similar configurations, \cite{Ge:2012bq} which studies the scattering matrix
\beq
{\underline {\underline S}}_0=\left[
\begin{array}{cc}
R_l & T \\
T & R_r 
\end{array}
\right].
\label{eq:SS}
\eeq
It can be shown that the eigenvalues $\sigma_1$ and $\sigma_2$ of such matrix are either both unimodular or of reciprocal magnitude,\cite{Ge:2012bq} viz.,
\beq
\left|
\sigma_1\sigma_2
\right|=1,
\label{eq:sigma12}
\eeq
with the two conditions $|\sigma_1|=|\sigma_2|=1$ and $|\sigma_1|=1/|\sigma_2|>1$ characterizing the so-called ``symmetric'' and ``broken'' phases, respectively.\cite{Ge:2012bq} The transition (usually referred to as ``exceptional point'') between these two phases was shown to be closely related to the transition (from real to conjugate-pairs) of the natural frequencies of the system in the complex frequency-plane, i.e., the onset of spontaneous symmetry breaking.\citep{Ge:2012bq} 
Accordingly, we monitor such transitions as a function of the frequency and incidence direction for a given level of loss and gain, or, equivalently, as a function of the gain/loss level for given frequency and incidence angle. 

Figure \ref{Figure10} illustrates this phenomenon, for three representative values of $\varepsilon''$ (corresponding to those chosen in Figs. \ref{Figure3}, \ref{Figure6}, and \ref{Figure8} above), by highlighting (cyan-shading) the regions in the parameter space $(d/\lambda_0,\theta)$ where the eigenvalues $\sigma_1$ and $\sigma_2$ of the scattering matrix ${\underline {\underline S}}_0$ in (\ref{eq:SS}) stay unimodular. It can be observed that, for a given value of $\varepsilon''$ and incidence angle, there exists a critical thickness value at which this transition occurs. For increasing values of $\varepsilon''$, this transition tends to occur at smaller values of $\theta$ and $d/\lambda_0$. It is also interesting to observe that, for incidence directions approaching the critical angle in (\ref{eq:N2lim}), the transition tends to occur at increasingly higher values of $d/\lambda_0$ [cf. Figs. \ref{Figure10}(a) and \ref{Figure10}(b)]. This is better quantified in Fig. \ref{Figure11}, which, for the case $\varepsilon''=0.001$, shows the magnitude of the two eigenvalues as a function of $d/\lambda_0$, for three incidence angles. It can be observed that, approaching the critical angle, the exceptional point moves towards increasingly larger values of $d/\lambda_0$.
This is not surprising since we have shown analytically that, in the asymptotic limit $k_0d\rightarrow\infty$ [also recalling (\ref{eq:unit2})],
\beq
R_l\rightarrow 0,~~R_r\rightarrow 0,~~\left|T\right|\rightarrow 1,
\eeq
which implies that the eigenvalues of scattering matrix in (\ref{eq:SS}) are both unimodular. In these conditions, no symmetry breaking occurs.

As previously mentioned, the tunneling conditions are generally different for the incidence from left and right, and hence
the phenomenon belongs to the general class of {\em anisotropic transmission resonances}.\cite{Ge:2012bq,Lin:2012} These phenomena, which feature
zero-reflection occurring only for incidence from one side of the structure and not from the other, have been observed in several in ${\cal PT}$-symmetric systems, and have been associated with exceptional points of the scattering matrix,\cite{Longhi:2011wy,Ge:2012bq}
\beq
{\underline {\underline S}}_c=\left[
\begin{array}{cc}
T & R_l \\
R_r & T 
\end{array}
\right],
\label{eq:SSc}
\eeq
which differs from that in (\ref{eq:SS}) by a mere permutation of its elements. The eigenvalues of this new scattering matrix ${\underline {\underline S}}_c$ (though different from those of ${\underline {\underline S}}_0$) exhibit the same property as in (\ref{eq:sigma12}). Moreover, while the matrix ${\underline {\underline S}}_0$ above exhibits a {\em single} transition,\citep{Ge:2012bq} the matrix ${\underline {\underline S}}_c$ may exhibit {\em multiple} exceptional points, which correspond to anisotropic transmission resonances. This is exemplified in Fig. \ref{Figure12}, which, for the case $\varepsilon''=0.001$ and $\theta=5^o$, shows the magnitude of the two eigenvalues as a function of $d/\lambda_0$. Two exceptional points can be observed, corresponding to the tunneling conditions for incidence from left and right (see also the magnified details in the inset).

\section{Conclusions and Outlook}
\label{Sec:Conclusions}
We have investigated tunneling phenomena that can occur in a ${\cal PT}$-symmetric ENZ bi-layer under obliquely-incident, TM-polarized plane-wave illumination. In particular, we have derived simple analytical conditions which parameterize the phenomenon, and also allow its physical interpretation in terms of the excitation of a surface-wave localized at the gain-loss interface. We have also identified a critical threshold of the level of gain and loss, below which the tunneling phenomenon may occur. Beyond a critical electrical thickness, the incidence direction at which the phenomenon occurs approaches a critical angle dictated by the surface-wave phase-matching condition.

Although the occurrence of this phenomenon is not strictly limited to the ENZ limit, in such regime much lower levels of loss and gain, and only moderately thick (wavelength sized) structures are required. 

Finally, via a spectral analysis, we have characterized the unidirectional character of this tunneling phenomenon, as well as the
 spontaneous symmetry breaking (i.e., transition from a real-valued to a complex-valued eigenspectrum) that can occur in this type of bi-layers.

The results from our prototype study constitute an interesting example of a tunneling phenomenon that is inherently induced by  ${\cal PT}$-symmetry, and may provide new perspectives in the effect of 
low levels of balanced loss and gain in ENZ materials. Current and future investigations are aimed at the study of more realistic configurations (based, e.g., on metal-dielectric multilayers), taking also into account the arising spatial dispersion.

\appendix 

\section{Details on the Derivation of Eqs. (\ref{eq:RLR}) and (\ref{eq:TLR})}
\label{Sec:AppA}
Assuming incidence from left, the $y$-directed magnetic field distribution in the various regions of Fig. \ref{Figure1} can be expressed as
\begin{widetext}
\beq
H_y\left(x,z\right)=\exp\left(i k_{x0}x\right)\left\{
\begin{array}{llll}
\exp\left[i k_{z0}\left(z+d\right)\right]+B_0\exp\left[-i k_{z0} \left(z+d\right)\right],\hspace{3mm}z<-d,\\
A_1 \exp\left(i k_{z1} z\right) +B_1\exp\left(-i k_{z1} z\right),\hspace{16mm} -d<z<0,\\
A_2 \exp\left(i k_{z1}^* z\right) +B_2\exp\left(-i k_{z1}^* z\right),\hspace{16mm} 0<z<d,\\
A_3 \exp\left[ik_{z0}\left(z-d\right)\right],\hspace{36mm}z>d,
\end{array}
\right.
\label{eq:ee}
\eeq
\end{widetext}
with $k_{x0}$ and $k_{z0}$ given in (\ref{eq:kxz0}), $k_{z1}$ given in (\ref{eq:kz1}), and the six unknown expansion coefficients $B_0, A_1, B_1, A_2, B_2, A_3$ to be calculated by enforcing the tangential-field continuity at the three interfaces $z=\pm d$ and $z=0$. Within this framework, the tangential electric field can be derived from (\ref{eq:ee}) and the appropriate Maxwell's curl equation, viz.,
\beq
E_x\left(x,z\right)=\frac{\eta_0}{ik_0\varepsilon\left(z\right)}\frac{\partial H_y}{\partial z}\left(x,z\right),
\eeq
with $\eta_0$ denoting the vacuum characteristic impedance, and
\beq
\varepsilon\left(z\right)=\left\{
\begin{array}{lll}
1,\hspace{6mm}\left|z\right|>d,\\
\varepsilon_1,\hspace{5mm} -d<z<0,\\
\varepsilon_1^*,\hspace{5mm} 0<z<d.
\end{array}
\right.
\eeq
In particular, we focus on the coefficients $B_0$ and $A_3$, which [cf. (\ref{eq:RRlr}) and (\ref{eq:TT})] play the role of the reflection and transmission coefficients. After cumbersome yet straightforward analytical manipulations, we obtain
\begin{subequations}
\beq
R_l=B_0\equiv \frac{N_{Rl}}{D},
\eeq
\begin{eqnarray}
N_{Rl}&=&\varepsilon_1 k_{z1} \tau_1^* \left[
\left(\varepsilon_1^*k_{z0}\right)^2-\left(k_{z1}^*\right)^2
\right]\nonumber\\
&+&\varepsilon_1^*k_{z1}^2\tau_1
\left(
i\varepsilon_1^*k_{z0}\tau_1^*-k_{z1}^*
\right)\nonumber\\
&+&\varepsilon_1^2k_{z0}k_{z1}^*\tau_1
\left(
\varepsilon_1^*k_{z0}+ik_{z1}^*\tau_1^*
\right),\\
D&=&\varepsilon_1^*k_{z1}^2\tau_1
\left(
k_{z1}^*-i\varepsilon_1^*k_{z0}\tau_1^*
\right)\nonumber\\
&+&\varepsilon_1^2k_{z0}k_{z1}^*\tau_1
\left(
\varepsilon_1^*k_{z0}-ik_{z1}^*\tau_1^*
\right)\nonumber\\
&+&\varepsilon_1k_{z1}
\left[
2i\varepsilon_1^*k_{z0} k_{z1}^*\right.
\nonumber\\
&+&
\left.
\left(\varepsilon_1^*\right)^2k_{z0}^2\tau_1^*+\left(
k_{z1}^*
\right)^2 \tau_1^*
\right],
\end{eqnarray}
\end{subequations}
\beq
T=A_3=\frac{2i k_{z0} \left|\varepsilon_1 \right|^2 \left|k_{z1}\right|^2 \sec\left(k_{z1}d\right) \sec\left(k_{z1}^*d\right)}{D}.
\eeq
The final results in (\ref{eq:RLR}) and (\ref{eq:TLR}) follow from further simplifications which exploit the ${\cal PT}$-symmetric character.

The reflection coefficient for incidence from right ($R_r$) can be computed by repeating the above analysis with the proper excitation or, more directly, by substituting $\varepsilon_1$ and $k_{z1}$ with their complex conjugates (and viceversa) in (\ref{eq:RLR}).

\section{Details on Eq. (\ref{eq:N13lim})}
\label{Sec:AppB}
First, we note from (\ref{eq:kz1}) and (\ref{eq:N2lim}) that
\beq
\bigl.k_{z1}\bigr|_{\theta=\theta_c}=\frac{\varepsilon_1 k_0}{\sqrt{2\varepsilon'}}.
\label{eq:kz1a}
\eeq
Moreover, it readily follows from (\ref{eq:tau1}) that
\beq
\lim_{k_0d\rightarrow\infty}\tau_1=-i.
\label{eq:tau1a}
\eeq
By substituting (\ref{eq:kz1a}) and (\ref{eq:tau1a}) in (\ref{eq:N1}), we obtain
\begin{eqnarray}
\lim_{k_0d\rightarrow\infty} N_1\left(\theta_c,k_0d,\varepsilon_1\right)&=&
\left|\varepsilon_1\right|^2 k_0 \left[
\frac{k_{z0}^2}{\sqrt{2\varepsilon'}}-\frac{k_0^2}{\left(2\varepsilon'\right)^{\frac{3}{2}}}
\right]\nonumber\\
&\times& \mbox{Re}\left(
i \left|\varepsilon_1\right|^2\right)=0,
\end{eqnarray}
which corresponds to (\ref{eq:N13lim}).

\section{Details on Eq. (\ref{eq:epsbound})}
\label{Sec:AppC}
In the ENZ limit (\ref{eq:ENZ}), we can neglect the term proportional to $\left|\varepsilon_1\right|^2$ in 
(\ref{eq:N1}), so that
\beq
N_1\left(\theta,k_0d,\varepsilon_1\right)\approx \left|k_{z1}\right|^2 \mbox{Re}\left(
\varepsilon_1^* k_{z1}\tau_1\right).
\label{eq:N12}
\eeq
In what follows, we determine the conditions under which a solution in $\theta$ of (\ref{eq:tunnel}) [with (\ref{eq:N12})] can be bracketed within the interval $(\theta_c,\pi/2)$. 

First, we note from (\ref{eq:N2}) and (\ref{eq:N2lim}) that the term $N_2$ vanishes for both $\theta=\theta_c$ and $\theta=\pi/2$.
Moreover, from (\ref{eq:tau1}) and (\ref{eq:kz1a}), we obtain
\beq
\bigl.\tau_1\bigr|_{\theta=\theta_c}= \tan\left(\frac{\varepsilon_1 k_0d}{\sqrt{2\varepsilon'}}\right),
\eeq
which, substituted in (\ref{eq:N12}) [with (\ref{eq:kz1a})], yields 
\beq
N_1\left(\theta_c,k_0d,\varepsilon_1\right)\approx\frac{\left|\varepsilon_1\right|^4k_0^3}{\left(2\varepsilon'\right)^{\frac{3}{2}}}
\mbox{Re}\left[
\tan\left(\frac{\varepsilon_1 k_0d}{\sqrt{2\varepsilon'}}\right)
\right].
\eeq
Recalling the behavior of the complex-argument tangent,\cite{Abramowitz:1964} we observe that the term $N_1$ is positive for
\beq
\mbox{Re}\left(\frac{\varepsilon_1 k_0d}{\sqrt{2\varepsilon'}}\right)\lesssim \frac{\pi}{2},
\eeq
which implies that
\beq
\frac{d}{\lambda_0}\lesssim \frac{1}{4}\sqrt{\frac{2}{\varepsilon'}}.
\label{eq:dl}
\eeq
We note that, in the ENZ limit $\varepsilon'\ll 1$, the condition in (\ref{eq:dl}) is verified for electrical thicknesses up to moderately large values. For instance, assuming $\varepsilon'=10^{-4}$, the condition is satisfied for $d/\lambda_0\lesssim 35$, i.e., well within the parameter range of interest. Therefore, we can conclude that the left-hand-side in (\ref{eq:N12}) is positive at $\theta=\theta_c$.

Next, we observe from (\ref{eq:kz1}) that
\beq
\bigl.k_{z1}\bigr|_{\theta=\frac{\pi}{2}}=k_0\sqrt{\varepsilon_1-1}\approx k_0\left(\frac{\varepsilon_1}{2}-i\right),
\label{eq:kz1b}
\eeq
where the approximate equality stems from a first-order McLaurin expansion (in $\varepsilon_1$). Similarly, by first-order McLaurin expansion (in $\varepsilon_1$) of $\tau_1$ in (\ref{eq:tau1}), we obtain 
\beq
\bigl.\tau_1\bigr|_{\theta=\frac{\pi}{2}}\approx -i\tau_0-i\frac{\varepsilon_1 k_0d}{2}\left(\tau_0^2-1\right),
\label{eq:tau1b}
\eeq
with $\tau_0$ defined in (\ref{eq:epsbound}). By substituting (\ref{eq:kz1b}) and (\ref{eq:tau1b}) in (\ref{eq:N12}), we obtain
\begin{eqnarray}
N_1\left(\frac{\pi}{2},k_0d,\varepsilon_1\right)&\approx&\frac{k^3_0\left|\varepsilon_1-1\right|}{4}\left[
2\tau_0 \left(\left|\varepsilon_1\right|^2-2\varepsilon'\right)
\right.\nonumber\\
&+&
\left.
\left|\varepsilon_1\right|^2 k_0 d\left(\varepsilon'-2\right)
\left(\tau_0^2-1\right)
\right].
\label{eq:N1pi2}
\end{eqnarray}
The condition in (\ref{eq:epsbound}) follows by enforcing that the expression in (\ref{eq:N1pi2}) is negative, so that a solution of (\ref{eq:N12}) can be bracketed within the interval $(\theta_c,\pi/2)$. We verified numerically that, within the parameter range of interest, $N_1$ is a monotonic function of $\theta$, and hence the above condition is not only sufficient, but also necessary.
 We stress that this result does not depend on the incidence side.




%

\end{document}